\documentclass[runningheads]{llncs}
\usepackage{booktabs}
\usepackage{multirow}
\usepackage{graphicx}
\usepackage{amsmath}
\usepackage{amssymb}
\usepackage{siunitx}
\usepackage{subcaption}
\usepackage{float}
\usepackage[misc,geometry]{ifsym}
\usepackage[colorlinks=true]{hyperref}

\usepackage{url}
\urldef{\mailsc}\path|haofu.liao@rochester.edu|

\usepackage{array}
\newcolumntype{C}[1]{>{\centering\let\newline\\\arraybackslash\hspace{0pt}}m{#1}}
\newcolumntype{?}{!{\vrule width 1pt}}

\begin{document}

\title{Artifact Disentanglement Network for Unsupervised Metal Artifact Reduction}
\titlerunning{Artifact Disentanglement Network}

\author{Haofu Liao\textsuperscript{1(\Letter)}
\and Wei-An Lin\textsuperscript{2}
\and Jianbo Yuan\textsuperscript{1}
\and S. Kevin Zhou\textsuperscript{3}
\and Jiebo Luo\textsuperscript{1}}
\authorrunning{H. Liao et al}

\institute{\textsuperscript{1} Department of Computer Science, University of Rochester\\ \mailsc\\
\textsuperscript{2}  Department of ECE, University of Maryland, College Park\\
\textsuperscript{3} Institute of Computing Technology, Chinese Academy of Sciences}%

\maketitle

\begin{abstract}

Current deep neural network based approaches to computed tomography (CT) metal artifact reduction (MAR) are supervised methods which rely heavily on synthesized data for training. 
However, as synthesized data may not perfectly simulate the underlying physical mechanisms of CT imaging, the supervised methods often generalize poorly to clinical applications. 
To address this problem, we propose, to the best of our knowledge, the first unsupervised learning approach to MAR. Specifically, we introduce a novel artifact disentanglement network that enables different forms of generations and regularizations between the artifact-affected and artifact-free image domains to support unsupervised learning. Extensive experiments show that our method significantly outperforms the existing unsupervised models for image-to-image translation problems, and achieves comparable performance to existing supervised models on a synthesized dataset. When applied to clinical datasets, our method achieves considerable improvements over the supervised models. The source code of this paper is publicly available at \url{https://github.com/liaohaofu/adn}.

\end{abstract}

\section{Introduction}

Metal artifact is one of the commonly encountered artifacts in computed tomography (CT) images. It is introduced by the metallic implants during the imaging and reconstruction process. The formation of metal artifact involves several mechanisms such as beam hardening, scatter, noise, and the non-linear partial volume effect \cite{mar_review}, which make it very challenging to be modeled and removed by traditional methods. Therefore, recent approaches \cite{cnnmartmi,cgan-mar,rl-arcnn,destreaknet,mpn,svar_gan,dudonet} to metal artifact reduction (MAR) propose to use deep neural networks (DNNs) to inherently address the modeling of metal artifacts, and their experimental results show promising MAR performances.

All the existing DNN-based approaches are supervised methods requiring pairs of anatomically identical CT images, one with and the other without metal artifacts, for training. As it is clinically impractical to obtain such pairs of images, most of the supervised methods rely on synthesized images to train their models. However, due to the complexity of metal artifacts and the variations of CT devices, the synthesized images may not fully simulate the real clinical scenarios, and the performances of these supervised methods may degrade in clinical applications.

In this work, we aim to address the challenging yet more practical unsupervised setting where {\it no paired CT images are available for training.} To this end, we propose a novel artifact disentanglement network to separate the metal artifacts from clinical CT images in a latent space. The disentanglement enables manipulations between the artifact-affected and artifact-free image domains so that different forms of adversarial- and self-regularizations can be achieved to support unsupervised learning. \textit{To the best of our knowledge, this is the first unsupervised learning approach to MAR}. Extensive experiments show that our method achieves comparable performance to the existing supervised methods on a synthesized dataset. When applied to clinical datasets, all the supervised methods demonstrate certain degrees of degradation, whereas our method outperforms the supervised methods with significantly better clinical MAR results.

\section{Related work}

{\noindent \bf Unsupervised image-to-image translation \hspace{5pt}} Image artifact reduction can be regarded as a form of image-to-image translation. One of the earliest unsupervised works in this category is CycleGAN \cite{cyclegancvpr} where a cycle-consistency design is proposed for unsupervised learning. Later works \cite{muniteccv,driteccv} improve CycleGAN for diverse and multimodal image generation. However, these unsupervised methods target at image synthesis and do not have suitable components for artifact reduction. Another recent work that is specialized for artifact reduction is deep image prior (DIP)~\cite{dipcvpr}, which, however, only works for less structured artifacts such as noise and compression artifacts.

{\vspace{0.5em} \noindent \bf Deep metal artifact reduction \hspace{5pt}} A number of studies have recently been proposed to address MAR with DNNs. RL-ARCNN \cite{rl-arcnn} introduces residual learning to a deep convolutional neural network (CNN) and achieves better MAR performance than ordinary CNN. DesteakNet \cite{destreaknet} proposes a two-streams approach that can take a pair of NMAR \cite{nmar} and detail images as the input to jointly reduce metal artifact. CNNMAR \cite{cnnmartmi} uses CNN to generate prior images in the CT image domain to help the correction in the sinogram domain. Both DesteakNet and CNNMAR show significant improvements over the existing non-DNN based methods on synthesized datasets. cGANMAR \cite{cgan-mar} leverages generative adversarial networks (GANs) \cite{gan} to further improve DNN-based MAR performance. 

\section{Methodology}

\begin{figure}[t]
\includegraphics[width=\textwidth]{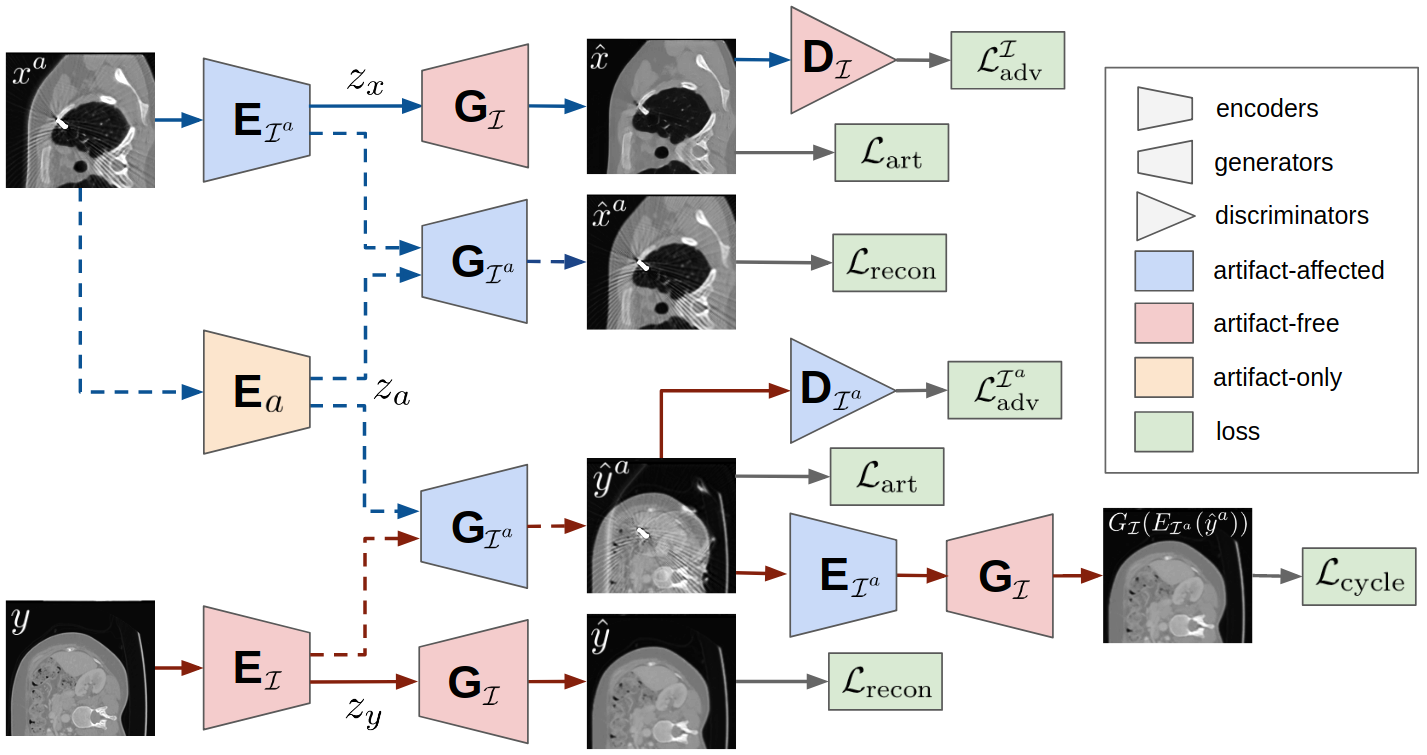}
\caption{Overview of the artifact disentanglement network.}
\label{fig:overview}
\end{figure}

Let $\mathcal{I}$ be the domain of all artifact-free CT images and $\mathcal{I}^{a}$ be the domain of all artifact-affected CT images, the proposed artifact disentanglement network (ADN) aims to learn a mapping from $\mathcal{I}^{a}$ to $\mathcal{I}$ without paired data. As illustrated in Figure \ref{fig:overview}, ADN contains a set of artifact-free image encoder, generator and discriminator $\{E_{\mathcal{I}}, G_{\mathcal{I}}, D_{\mathcal{I}} \}$, a set of artifact-affected image encoder, generator and discriminator $\{E_{\mathcal{I}^{a}}, G_{\mathcal{I}^{a}}, D_{\mathcal{I}^{a}}\}$ and an artifact-only encoder $E_{a}$. The architectures of these building components are inspired from the state-of-the-art studies for image-to-image translation \cite{cyclegan,muniteccv}. See the supplementary material for their detailed structures.

{\vspace{0.5em} \noindent \bf Components \hspace{5pt}} Given two unpaired images $x^a \in \mathcal{I}^{a} $ and $ y \in \mathcal{I}$, the encoders $E_{\mathcal{I}^{a}}$ and $E_{\mathcal{I}}$ map the artifact-free 
content information from $x^a$ and $y$ to a common content space $\mathcal{C}$, respectively. $E_{a}$ maps the artifact-only information from $x^a$ to an artifact space $\mathcal{A}$,
\begin{equation}
     z_x = E_{\mathcal{I}^a}(x^a), z_y = E_{\mathcal{I}}(y), z_a = E_{a}(x^a), \quad \{z_x, z_y\} \subset \mathcal{C}, z_a \in \mathcal{A}.
\end{equation}
The generator $G_{\mathcal{I}^{a}}$ takes an artifact-free code, $z_x$ or $z_y$, and an artifact-only code $z_a$ as the input and outputs an artifact-affected image. $G_{\mathcal{I}}$ takes an artifact-free code, $z_x$ or $z_y$, as the input and outputs an artifact-free image,
\begin{equation}
\begin{split}
\hat{x} = G_{\mathcal{I}}(z_x),& \quad \hat{x}^a = G_{\mathcal{I}^a}(z_x, z_a), \\
\quad \hat{y} = G_{\mathcal{I}}(z_y),& \quad \hat{y}^a = G_{\mathcal{I}^a}(z_y, z_a).\\
\end{split}
\end{equation}
During testing, only $E_{\mathcal{I}^{a}}$ and $G_{\mathcal{I}}$ are required to obtain an artifact-corrected output, i.e., $\hat{x} = G_{\mathcal{I}}(E_{\mathcal{I}^{a}}(x^a))$. The discriminator $D_{\mathcal{I}^{a}}$ decides whether an input is sampled from $\mathcal{I}^{a}$ or generated by $G_{\mathcal{I}^{a}}$. Similarly, $D_{\mathcal{I}}$ decides whether an input is from $\mathcal{I}$ or $G_{\mathcal{I}}$.

{\vspace{0.5em} \noindent \bf Loss functions \hspace{5pt}} A good MAR model should (i) reduce the artifacts as much as possible and (ii) keep the anatomical content of the input CT images. To remove the artifacts, we train  $D_{\mathcal{I}}$ and $G_{\mathcal{I}}$ adversarially to encourage the 
output $\hat{x}$ to appear similar to an artifact-free image,
\begin{equation}
\mathcal{L}_{\text{adv}}^{\mathcal{I}} = \mathbb{E}_{\mathcal{I}}[\log D_{\mathcal{I}}(y)]
 + \mathbb{E}_{\mathcal{I}^{a}}[1 - \log D_{\mathcal{I}}(\hat{x})]
\end{equation}
To maintain the anatomical content, we apply self-reconstruction to force the encoders and decoders to preserve the content of the inputs,
\begin{equation}
    \mathcal{L}_{\text{recon}} = \mathbb{E}_{\mathcal{I},\mathcal{I}^a}[||\hat{x}^a - x^a||_1 +||\hat{y} - y||_1].
\end{equation}
Here, the first term encourages $E_{\mathcal{I}^{a}}$ encodes all the content information of $x^a$ and the artifact information is not encoded due to the introduction of a separate artifact encoder $E_{a}$. With the second term, $G_{\mathcal{I}}$ learns how to fully reconstruct the encoded artifact-free content information. Combining these two terms, content persevering for $\hat{x}$ can be achieved.

In addition, we also introduce a {\it self-reduction design} to further enforce the learning. This idea is carried out in two steps. In the first step, ADN synthesizes ``real'' metal artifact from $x^a$ and apply it to $y$. Specifically, this is achieved by decoding from $z_y$ and $z_a$, i.e., $\hat{y}^a = G_{\mathcal{I}^a}(z_y, z_a)$, and we use another adversarial loss to guarantee $\hat{y}^a$ looking ``real'',
\begin{equation}
    \mathcal{L}_{\text{adv}}^{\mathcal{I}^{a}} = \mathbb{E}_{\mathcal{I}^a}[\log D_{\mathcal{I}^a}(x^a)] + \mathbb{E}_{\mathcal{I},\mathcal{I}^a}[1 - \log D_{\mathcal{I}^a}(\hat{y}^a)]
\end{equation}
In the second step, ADN reduces artifacts from the synthesized data to recover back to $y$. This is regularized by a cycle-consistent loss
\begin{equation}
    \mathcal{L}_{\text{cycle}} = \mathbb{E}_{\mathcal{I},\mathcal{I}^a}[|| G_{\mathcal{I}}(E_{\mathcal{I}^{a}}(\hat{y}^a)) - y||_1].
\end{equation}

Finally, due to the use of the same metal artifact, the difference map between $x^a$ and $\hat{x}$ and 
that between $\hat{y}^a$ and $y$ should be close. Thus, we employ an {\it artifact-consistent} loss to constrain the artifact difference,
\begin{equation}
    \mathcal{L}_{\text{art}} = \mathbb{E}_{\mathcal{I},\mathcal{I}^a}[||(x^a - \hat{x}) - (\hat{y}^a - y)||_1].
\end{equation}
The full objective function is given by
\begin{equation}
    \mathcal{L} = \lambda_{\text{adv}}^{\mathcal{I}} \mathcal{L}_{\text{adv}}^{\mathcal{I}} + \lambda_{\text{adv}}^{\mathcal{I}^a} \mathcal{L}_{\text{adv}}^{\mathcal{I}^{a}} + \lambda_{\text{recon}} \mathcal{L}_{\text{recon}} + \lambda_{\text{cycle}} \mathcal{L}_{\text{cycle}} + \lambda_{\text{art}} \mathcal{L}_{\text{art}},
\end{equation}
where the $\lambda$'s are hyper-parameters that control the importance of each term.

\section{Experiments}

{\vspace{0.5em} \noindent \bf Datasets. \hspace{5pt}} We evaluate the proposed method on one synthesized dataset and two clinical datasets. We refer to them as SYN, CL1 and CL2, respectively. For SYN, we randomly select $4,118$ artifact-free CT images from DeepLesion \cite{deeplesion} and follow the method from CNNMAR \cite{cnnmartmi} to synthesize metal artifacts. We use $3,918$ of the synthesized pairs for training and validation and the rest $200$ pairs for testing.

For CL1, we choose the vertebrae localization and identification dataset from Spineweb\footnote{spineweb.digitalimaginggroup.ca}.
We split the CT images from this dataset into two groups, one with artifacts and the other without artifacts. First, we identify regions with HU values greater than $2,500$ as the metal regions. Then, CT images whose largest-connected metal regions have more than 400 pixels are selected as artifact-affected images. CT images with the largest HU values less than $2,000$ are selected as artifact-free images. After this selection, the artifact-affected group contains $6,270$ images and the artifact-free group contains $21,190$ images. We withhold $200$ images from the artifact-affected group for testing.

For CL2, we investigate the performance of the proposed method under a more challenging {\it cross-modality} setting. Specifically, the artifact-affected images of CL2 are from a cone-beam CT (CBCT) dataset collected during spinal interventions. Images from this dataset are very noisy and the majority of them contain metallic implants. There are in total $2,560$ CBCT images from this dataset, among which 200 images are withheld for testing. For the artifact-free images, we reuse the CT images collected from CL1.

\begin{table}[t]
\centering
\caption{Quantitative evaluation on the SYN dataset.}
\begin{tabular}{@{}c?ccc?ccccc@{}}
\noalign{\hrule height 1pt}
 \multirow{2}{*}{} & \multicolumn{3}{c?}{\bf Supervised} & \multicolumn{5}{c}{\bf Unsupervised}               \\  
       &\tiny{\bf CNNMAR}\cite{cnnmartmi}    & \scriptsize{\bf UNet}  \cite{unet} & \tiny{\bf cGANMAR \cite{cgan-mar}} & \scriptsize{\bf Ours} & \tiny{\bf CycleGAN} \cite{cgan-mar} & \scriptsize{\bf DIP} \cite{dipcvpr} & \tiny{\bf MUNIT} \cite{muniteccv} & \scriptsize{\bf DRIT} \cite{driteccv} \\ \noalign{\hrule height 1pt}
\scriptsize{\bf PSNR} & 32.5 & \bf 34.8              & \bf 34.1           & \underline{33.6}      & 30.8     & 26.4 & 14.9  & 25.6 \\
\scriptsize{\bf SSIM} & 91.4 & \bf 93.1              & \bf 93.4           & \underline{92.4}      & 72.9     & 75.9 & 7.5   & 79.7 \\
\noalign{\hrule height 1pt}
\end{tabular}%
\label{tab:metrics}
\end{table}

\begin{figure}[t]
\includegraphics[width=\textwidth]{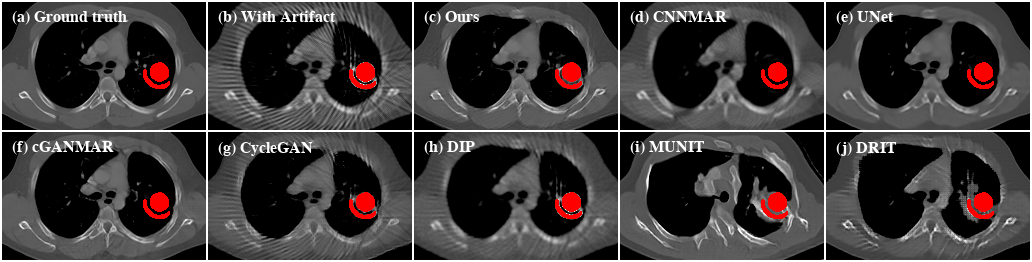}
\caption{Qualitative evaluation on the SYN dataset. For better visualization, we obtain the metal region through thresholding and color it with red. See the supplementary material for more qualitative results.}
\label{fig:syn}
\end{figure}

{\vspace{0.5em} \noindent \bf Baselines. \hspace{5pt}} We compare the proposed method with seven state-of-the-art methods that are closely related to our problem. Three of the compared methods are supervised: CNNMAR \cite{cnnmartmi}, UNet \cite{unet} and cGANMAR \cite{cgan-mar}. CNNMAR and cGANMAR are two recent approaches that are dedicated to MAR. UNet is a general DNN framework that shows effectiveness in many image-to-image problems. The other four compared methods are unsupervised: CycleGAN \cite{cyclegancvpr}, DIP \cite{dipcvpr}, MUNIT \cite{muniteccv} and DRIT \cite{driteccv}. These methods are currently state-of-the-art approaches to unsupervised image-to-image translation problems.
All the compared methods except UNet are trained with their officially released code. For UNet, a publicly available implementation\footnote{github.com/milesial/Pytorch-UNet} is used.

{\vspace{0.5em} \noindent \bf Training and testing. \hspace{5pt}} We implement our method under the PyTorch deep learning framework\footnote{pytorch.org} and use the Adam optimizer with $\num{1e-4}$ learning rate to minimize the objective function. For the hyper-parameters, we use $\lambda_{\text{adv}}^{\mathcal{I}}=\lambda_{\text{adv}}^{\mathcal{I}^a}=1.0$, $\lambda_{\text{recon}}=\lambda_{\text{cycle}}=\lambda_{\text{art}}=20.0$ for SYN and CL1, and use $\lambda_{\text{adv}}^{\mathcal{I}}=\lambda_{\text{adv}}^{\mathcal{I}^a}=1.0$, $\lambda_{\text{recon}}=\lambda_{\text{cycle}}=\lambda_{\text{art}}=5.0$ for CL2.

To simulate the unsupervised setting for SYN, we evenly divide the $3,918$ synthesized training pairs into two groups. For one group, only artifact-affected images are used and their corresponding artifact-free images are withheld. For the other group, only artifact-free images are used and their corresponding artifact-affected images are withheld. During training of the unsupervised methods, we randomly select one image from each of the two groups as the input. For the supervised methods, all the $3,918$ synthesized training pairs are used.

To train the supervised methods with CL1, we first synthesize metal artifacts using the images from the artifact-free group of CL1. Then, we train the supervised methods with the synthesized pairs. During testing, the trained models are applied to the testing set containing only clinical metal artifact images. To train the unsupervised methods, we randomly select one image from the artifact-affected group and the other from the artifact-free group as the input.

For CL2, synthesizing metal artifacts is not possible due to the unavailability of artifact-free CBCT images. Therefore, for the supervised methods we directly use the models trained for CL1. In other words, the supervised methods are trained on synthesized CT images (from CL1) and tested on clinical CBCT images (from CL2). For the unsupervised models, each time we randomly select one artifact-affected CBCT image and one artifact-free CT image as the input for training.

\begin{figure}[t]
\includegraphics[width=\textwidth]{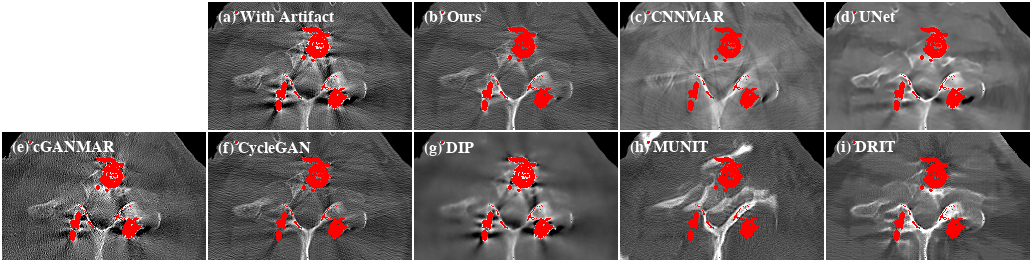}
\caption{Qualitative evaluation on the CL1 dataset. For better visualization, we obtain the metal region through thresholding and color it with red. See the supplementary material for more qualitative results.}
\label{fig:cl1}
\end{figure}

{\vspace{0.5em} \noindent \bf Performance on synthesized data.  \hspace{5pt}} SYN contains paired data, allowing for both quantitative and qualitative evaluations. Following the convention in the literature, we use peak signal-to-noise ratio (PSNR) and structural similarity index (SSIM) as the metrics for the quantitative evaluation. For both metrics, the higher the better. Table \ref{tab:metrics} and Figure \ref{fig:syn} show the quantitative and qualitative evaluation results, respectively.

We observe that the proposed method performs significantly better than the other unsupervised methods. MUNIT focuses more on diverse and realistic outputs (Figure \ref{fig:syn}(i)) with less constraint on structural similarity. CycleGAN and DRIT perform better as both the two models also require the artifact-corrected outputs to be able to transform back to the original artifact-affected images. Although this helps preserve content information, it also encourages the models to keep the artifacts. Therefore, as shown in Figure \ref{fig:syn}(g) and 2(j), the artifacts cannot be greatly reduced. DIP does not reduce much metal artifact in the input image (Figure \ref{fig:syn}(h)) as it is not designed to handle the more structured metal artifact.

We also find that the performance of our method is on a par with the supervised methods. The performance of UNet 
is close to that of cGANMAR which at its backend uses an UNet-like architecture. However, owing to the use of GAN, it produces sharper outputs (Figure \ref{fig:syn}(e)) than UNet (Figure \ref{fig:syn}(f)). As for PSNR and SSIM, both methods only slightly outperform our method and, surprisingly, our method performs better than CNNMAR.

{\vspace{0.5em} \noindent \bf Performance on clinical data.  \hspace{5pt}} Next, we investigate the performance of the proposed method on clinical data. Since there are no ground truths available for the clinical images, only qualitative comparisons are performed. The qualitative evaluation results of CL1 are shown in Figure \ref{fig:cl1}. Here, all the supervised methods are trained with paired images that are synthesized from the artifact-free group of CL1. We can see that UNet and cGANMAR generalize poorly when applied to clinical images (Figure \ref{fig:cl1}(d) and \ref{fig:cl1}(e)). CNNMAR is more robust as it corrects the artifacts in the sinogram domain. However, such a sinogram domain correction also introduces secondary artifacts (Figure \ref{fig:cl1}(c)). For the more challenging cross-modality artifact reduction task with CL2 (Figure \ref{fig:cl2}), all the supervised methods fail. This is not totally unexpected as the supervised methods are trained using only CT images because of the lack of artifact-free CBCT images. Similar to the cases with SYN, the other unsupervised methods also show inferior performances when evaluated on both the CL1 and CL2 datasets. By contrast, our method consistently delivers high-quality artifact reduced results on clinical images.

\begin{figure}[t]
\includegraphics[width=\textwidth]{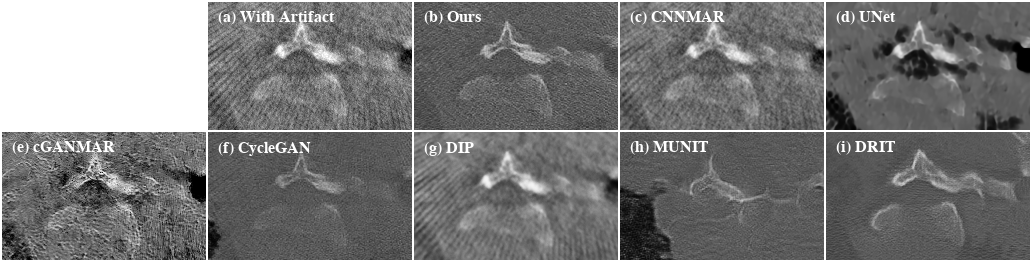}
\caption{Qualitative evaluation on the CL2 dataset. See the supplementary material for more qualitative results.}
\label{fig:cl2}
\end{figure}

\section{Conclusion}

We presented a novel unsupervised learning approach to MAR. Through the development of an artifact disentanglement network, we showed how to leverage different forms of regularizations to eliminate the requirement of paired images for training. To understand the effectiveness of this approach, we performed extensive evaluations on one synthesized and two clinical datasets. The evaluation results demonstrated the feasibility of using unsupervised learning method to achieve comparable performance to the supervised methods. More importantly, the results also showed that directly learning MAR from clinical CT images under an unsupervised setting was a more feasible and robust approach than transferring the knowledge learned from synthesized data to clinical data. We believe our findings in this work will initiate more applicable research for medical image artifact reduction even under an unsupervised setting. \newline

\noindent \textbf{Acknowledgement.} This work was supported in part by NSF award \#1722847 and the Morris K. Udall Center of Excellence in Parkinson's Disease Research by NIH.

\bibliographystyle{splncs04}
\bibliography{references}

\newpage

\noindent \vspace{.1em}
\begin{center}
\textbf{\LARGE{Supplementary Material}}\vspace{3em}
\end{center}

\renewcommand\thesection{\Alph{section}}
\setcounter{section}{0}

\section{Architecture Details}

\begin{figure}
\centering
\includegraphics[width=0.7\textwidth]{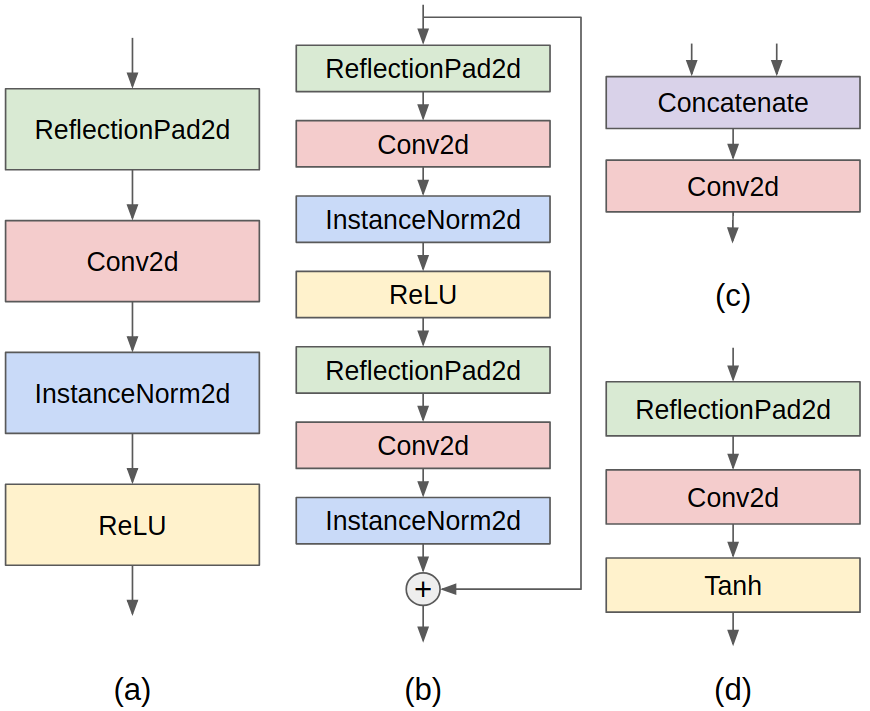}
\caption{Basic building blocks of the encoders and generators: (a) convolution block, (b) residual block, (c) merge block, and (d) final block. ReflectionPad2d stands for a reflection padding layer that we use to replace the zero padding of the conventional convolution layer.}
\label{fig:blocks}
\end{figure}

\begin{figure}
\centering
\includegraphics[width=0.6\textwidth]{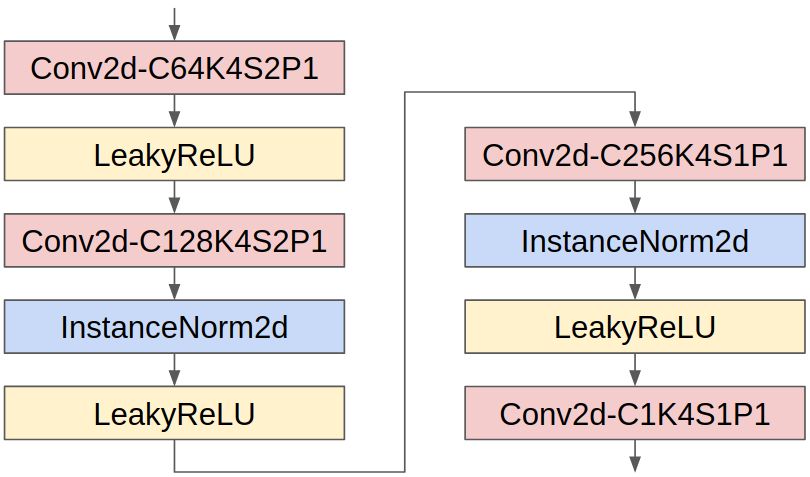}
\caption{Architecture of the discriminator $D_{\mathcal{I}}$ or $D_{\mathcal{I}^a}$.  We use `C\#K\#S\#P\#' to denote the configuration of the convolution layers, where `K', `C', `S' and `P' stand for the kernel, output channel, stride and padding size, respectively.}
\label{fig:discriminator}
\end{figure}

\begin{figure}
\centering
\includegraphics[width=\textwidth]{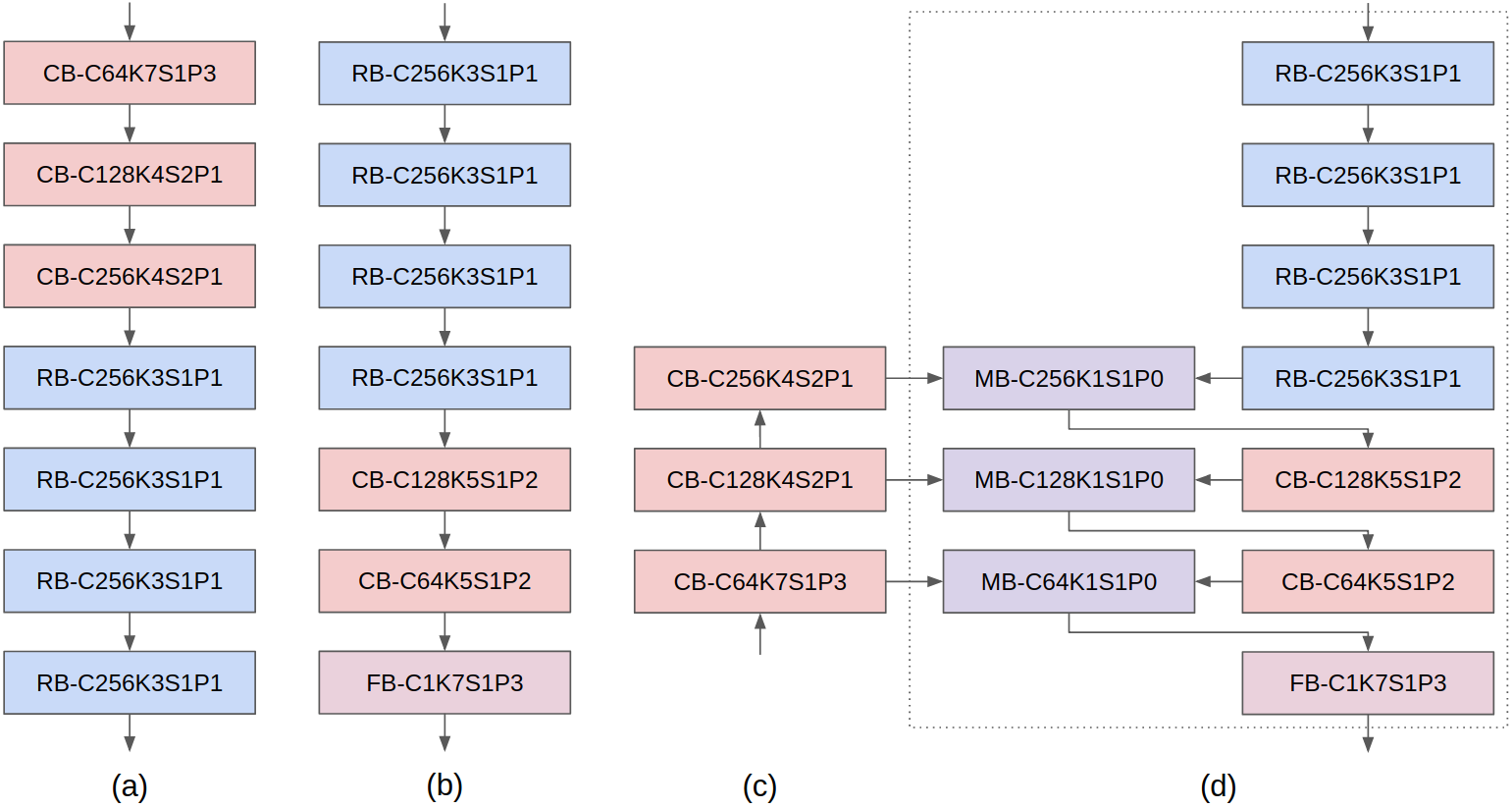}
\caption{Architecture of the encoders and generators. (a) $E_{\mathcal{I}}$ or  $E_{\mathcal{I}^a}$ (b) $G_{\mathcal{I}}$ (c) $E_{a}$ (d) $G_{\mathcal{I}^a}$. CB, RB, MB and FB are acronyms of the build blocks as illustrated in Fig. \ref{fig:blocks}. The same as in Fig. \ref{fig:discriminator}, `C\#K\#S\#P\#' denotes the configurations of the convolution layers in the blocks. For CB, RB, and FB, P is the padding of the reflection padding layer and the padding of the convolutional layer is zero. Note that the artifact code input for $G_{\mathcal{I}^a}$ are the hierarchical features encoded by $E_{a}$ and are merged with the corresponding outputs from $G_{\mathcal{I}^a}$.}
\end{figure}

\newpage
\section{Qualitative Results}
\begin{figure}[H]
\centering
\begin{subfigure}{0.98\textwidth}
  \centering
  \includegraphics[width=\linewidth]{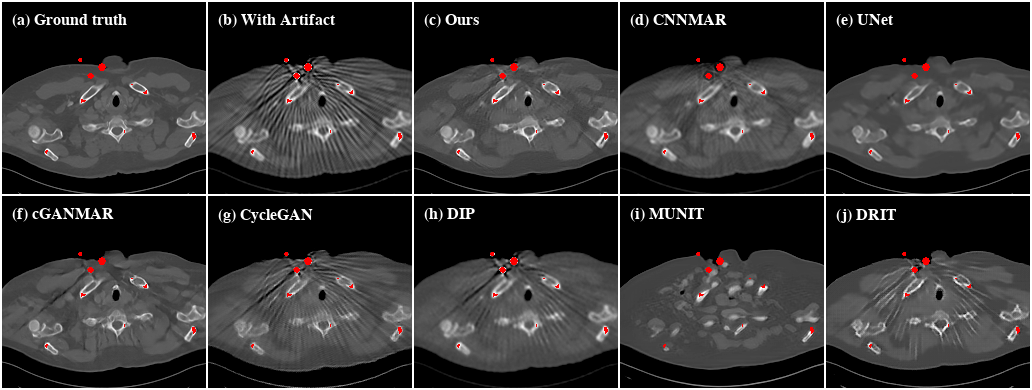}
\end{subfigure}\\
\vspace{.3em}
\begin{subfigure}{0.98\textwidth}
  \centering
  \includegraphics[width=\linewidth]{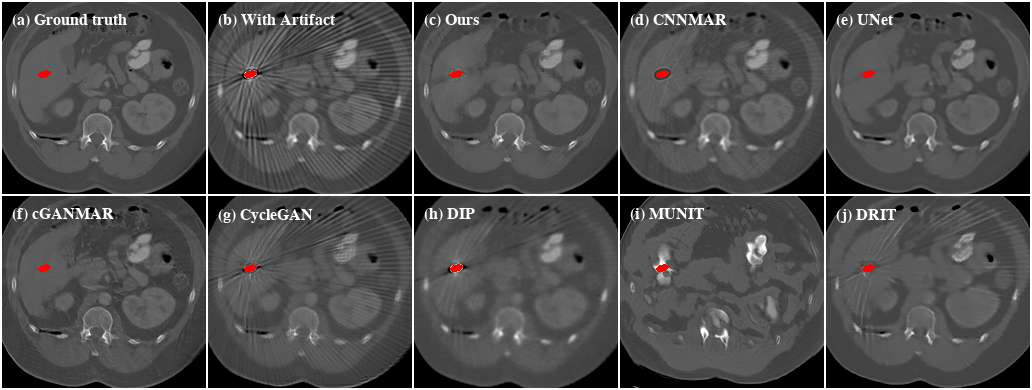}
\end{subfigure}\\
\vspace{.3em}
\begin{subfigure}{0.98\textwidth}
  \centering
  \includegraphics[width=\linewidth]{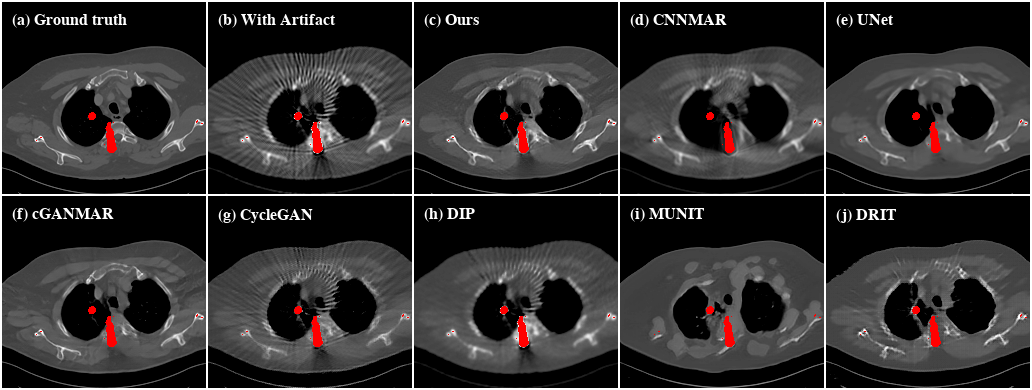}
\end{subfigure}\\
\vspace{.3em}
\begin{subfigure}{0.98\textwidth}
  \centering
  \includegraphics[width=\linewidth]{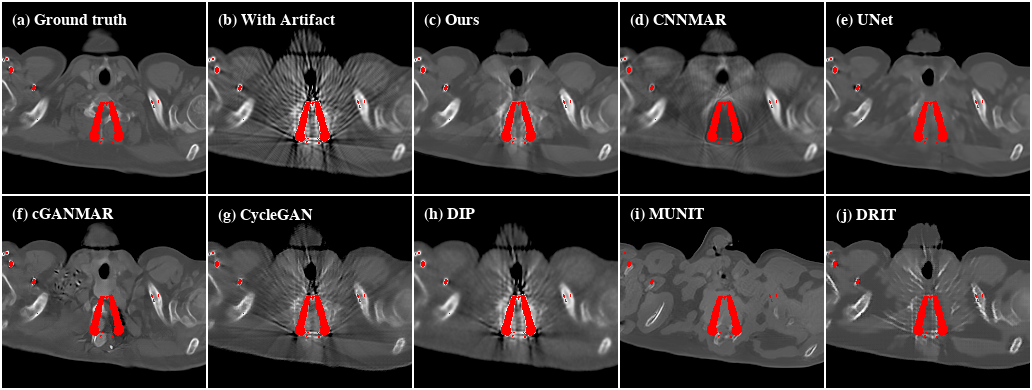}
\end{subfigure}
\caption{Qualitative evaluation results of SYN. For better visualization, we obtain the metal regions through thresholding and color them with red.}
\label{fig:syn}
\end{figure}

\begin{figure}[H]
\centering
\begin{subfigure}{0.98\textwidth}
  \centering
  \includegraphics[width=\linewidth]{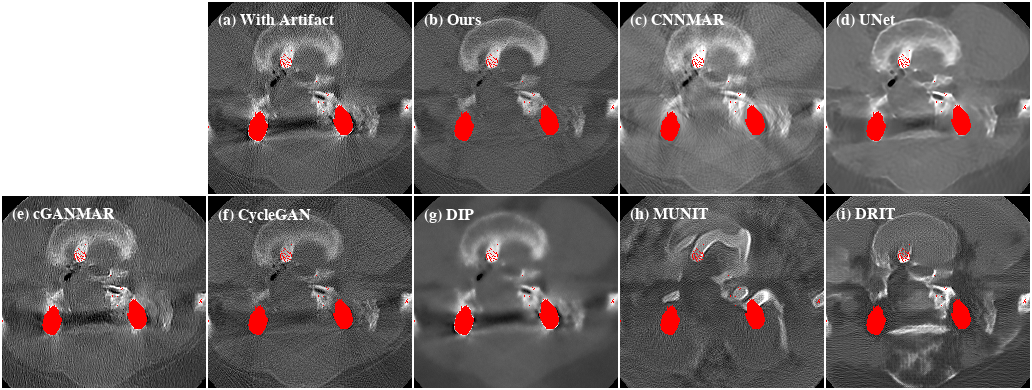}
\end{subfigure}\\
\vspace{.3em}
\begin{subfigure}{0.98\textwidth}
  \centering
  \includegraphics[width=\linewidth]{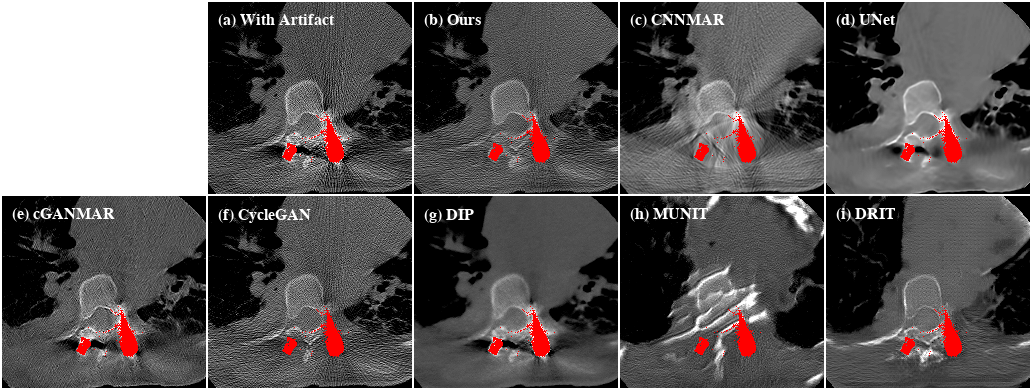}
\end{subfigure}\\
\vspace{.3em}
\begin{subfigure}{0.98\textwidth}
  \centering
  \includegraphics[width=\linewidth]{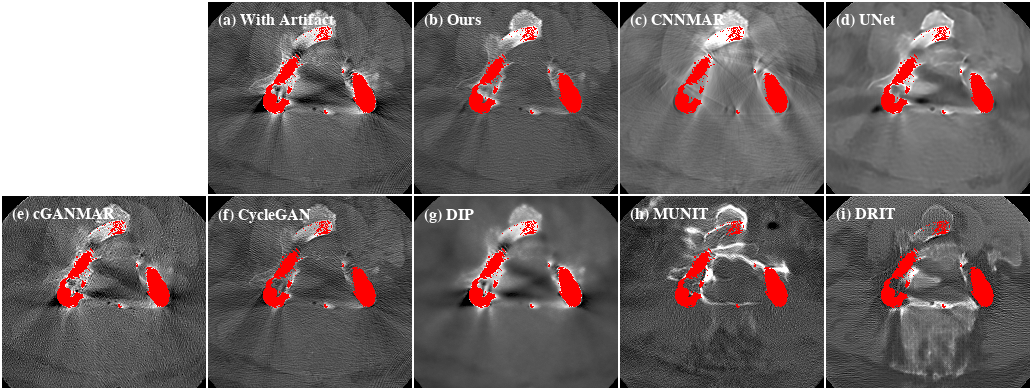}
\end{subfigure}\\
\vspace{.3em}
\begin{subfigure}{0.98\textwidth}
  \centering
  \includegraphics[width=\linewidth]{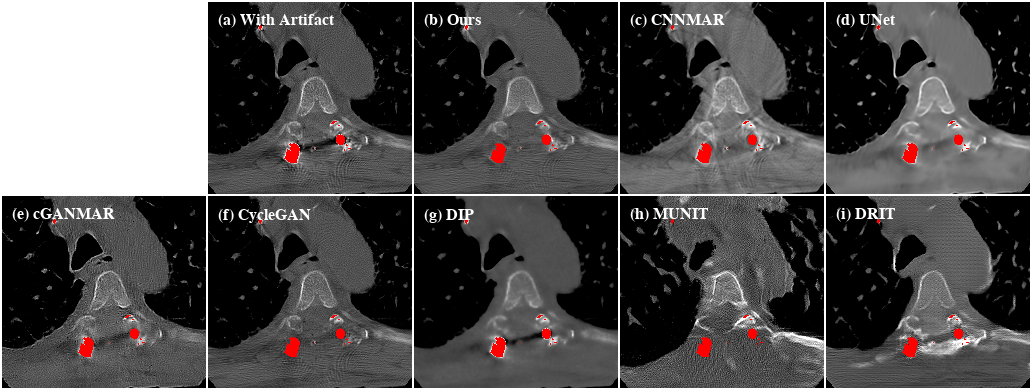}
\end{subfigure}
\caption{Qualitative evaluation results of CL1. For better visualization, we obtain the metal regions through thresholding and color them with red.}
\label{fig:cl1}
\end{figure}

\begin{figure}[H]
\centering
\begin{subfigure}{0.98\textwidth}
  \centering
  \includegraphics[width=\linewidth]{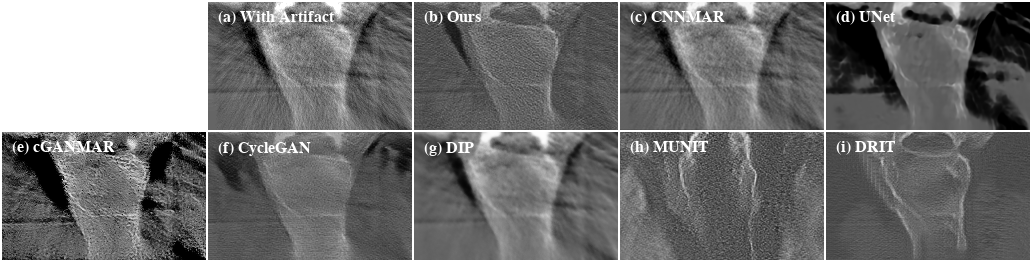}
\end{subfigure}\\
\vspace{.3em}
\begin{subfigure}{0.98\textwidth}
  \centering
  \includegraphics[width=\linewidth]{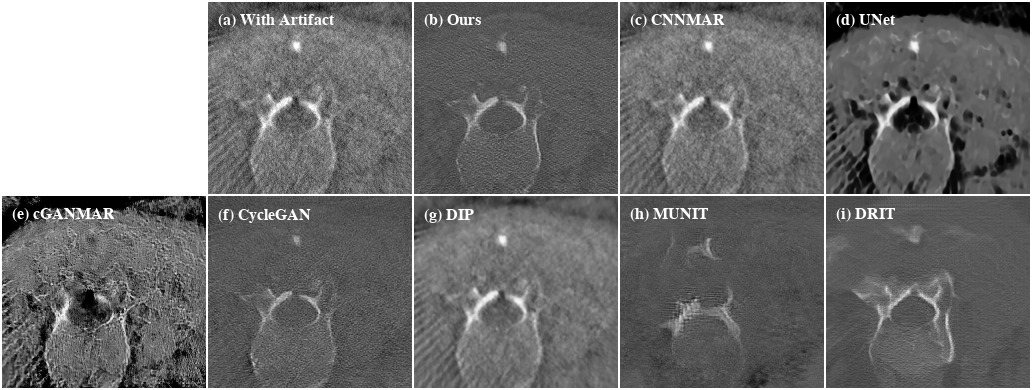}
\end{subfigure}\\
\vspace{.3em}
\begin{subfigure}{0.98\textwidth}
  \centering
  \includegraphics[width=\linewidth]{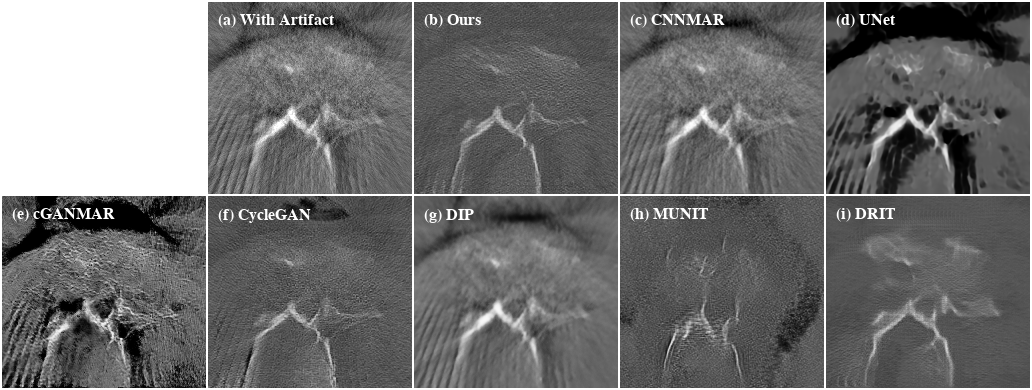}
\end{subfigure}\\
\vspace{.3em}
\begin{subfigure}{0.98\textwidth}
  \centering
  \includegraphics[width=\linewidth]{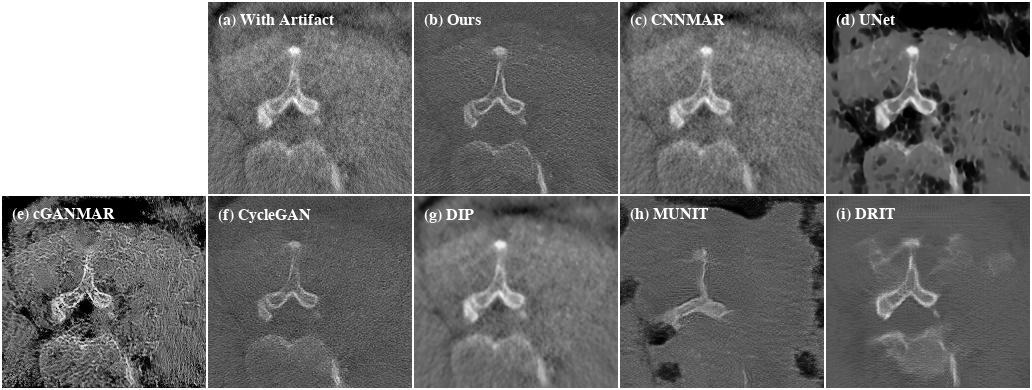}
\end{subfigure}
\caption{Qualitative evaluation results of CL2.}
\label{fig:cl2}
\end{figure}

\begin{figure}
\centering
\includegraphics[width=\textwidth]{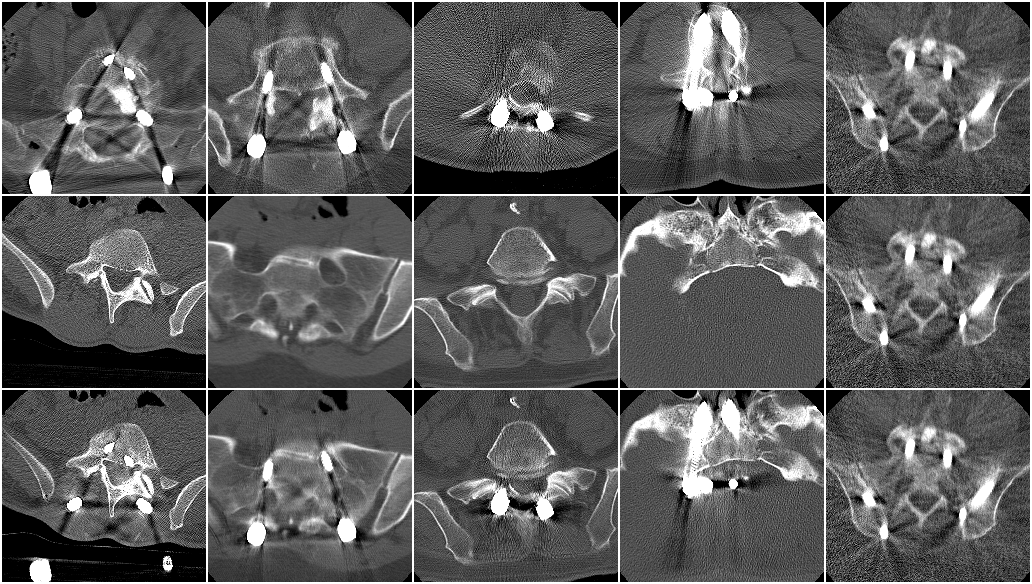}
\caption{Metal artifact transferring. First row: the clinical images with metal artifacts. Middle row: the clinical images without metal artifacts. Last row: the metal artifacts in the first row transferred to the artifact-free images in the second row. }
\end{figure}

\end{document}